\newif\ifstandardtemplate \standardtemplatetrue
\newif\ifelseviertemplate \elseviertemplatefalse
\newif\ifspringertemplate \springertemplatefalse
\newif\ifwileytemplate    \wileytemplatefalse

\newcommand{\mytitle}{Structural models based on 3D constitutive laws: variational
  structure and numerical solution}

\newcommand{\myabstract}{In all structural models, the
section or fiber response is a relation between the strain measures and the stress resultants. This relation can only be expressed in a simple analytical form when the material response is linear elastic. For
other, more complex and interesting situations, kinematic and kinetic hypotheses need to be invoked,
and a constrained three-dimensional constitutive relation has to be employed at every point of the
section in order to implement non-linear and dissipative constitutive laws into dimensionally reduced structural models. In this article we explain in which sense reduced constitutive models can be
expressed as minimization problems, helping to formulate the global equilibrium as a single
optimization problem. Casting the problem this way has implications from the mathematical and
numerical points of view, naturally defining error indicators.  General purpose solution algorithms
for constrained material response, with and without optimization character, are discussed and
provided in an open-source library.}

\newcommand{\mypackages}{
  \usepackage{graphicx}
  \graphicspath{{Figures/}{figures/}}
}

\ifspringertemplate
\title{the title%
\thanks{the thanks}}
\author{\ldots \and Ignacio Romero \and \ldots}

\institute{I. Romero \at
  IMDEA Materials Institute, Eric Kandel 2, Tecnogetafe, Madrid 28906, Spain\\
  Universidad Polit\'ecnica de Madrid, Jos\'e Guti\'errez Abascal, 2, Madrid 29006, Spain\\
  \email{ignacio.romero@imdea.org}
  \and
  XX \at
  XXX\\
  \email{XXX}}

\date{Received: date / Accepted: date}

\maketitle

\begin{abstract}
\myabstract 
\end{abstract}
\fi

\ifelseviertemplate
\documentclass[preprint,11pt]{elsarticle}
\mypackages
\newcommand{\mybibstyle}{elsarticle-num}
\usepackage{IBB_kurz}
\usepackage{gnuplot-lua-tikz}

\journal{Computer Methods in Applied Mechanics and Engineering}
\begin{document}
\begin{frontmatter}

\title{\mytitle}

\author[1,2]{David Portillo}
\ead{david.portillo@upm.es}

\author[3]{Bastian Oesterle}
\ead{oesterle@ibb.uni-stuttgart.de}

\author[3]{Rebecca Thierer}
\ead{thierer@ibb.uni-stuttgart.de}

\author[3]{Manfred Bischoff}
\ead{bischoff@ibb.uni-stuttgart.de}

\author[1,2]{Ignacio Romero\corref{cor1}}
\ead{ignacio.romero@upm.es}


\address[1]{Universidad Polit\'ecnica de Madrid, ETSI Industriales, Spain}
\address[2]{IMDEA Materials Institute, Spain}
\address[3]{University of Stuttgart, Institute for Structural Mechanics, Germany}

\cortext[cor1]{Corresponding author. ETS Ingenieros Industriales,
  Jos\'e Guti\'errez Abascal, 2, Madrid 28006, Spain}

\begin{abstract}
\myabstract
\end{abstract}

\end{frontmatter}

\fi

\ifstandardtemplate
\documentclass[10pt,a4paper]{article}
\usepackage{IBB_kurz}
\usepackage{gnuplot-lua-tikz}
\newcommand{\mybibstyle}{unsrt}

\message{Compiling with default template}

\title{\mytitle}
\author{D. Portillo$^{1,2}$, B. Oesterle$^3$, R. Thierer$^3$, M. Bischoff$^3$, I. Romero$^{1,2}$}
\date{$^1$Universidad Polit\'ecnica de Madrid, 
         Jos\'{e} Guti\'{e}rrez Abascal, 2, 28006 Madrid, Spain\\[2ex]%
         $^2$IMDEA Materials Institute, Eric Kandel 2, Tecnogetafe, Madrid 28906, Spain\\[2ex]%
         $^3$University of Stuttgart, Institute for Structural Mechanics, Pfaffenwaldring 7, 70550 Stuttgart, Germany\\[2ex]%
      \today}

\begin{document}
\maketitle
\fi

\newcommand{\concept}[1]{\textbf{\emph{#1}}}
\newcommand{\defined}{:=}
\newcommand{\mbs}[1]{\boldsymbol{#1}}
\newcommand{\dd}[2]{\frac{\mathrm{d} #1}{\mathrm{d} #2}}
\newcommand{\pd}[2]{\frac{\partial #1}{\partial #2}}
\newcommand{\fd}[2]{\frac{\delta #1}{\delta #2}}
\newcommand{\set}[1]{\left\{#1\right\}}
\newcommand{\trace}{{\mathop{\mathrm{tr}}}}

\newcommand{\mbb}[1]{\mathbb{#1}}
\newcommand{\mbf}[1]{\mathbf{#1}}
\newcommand{\dddd}[2]{\frac{\mathrm{d}^2\, #1}{\mathrm{d}\, #2^2}}

\newcommand{\first}[1]{\bar{#1}}
\newcommand{\second}[1]{\tilde{#1}}
\newcommand{\ke}{{\mbs{\varepsilon}_1}}
\newcommand{\ue}{\mbs{\varepsilon}_2}


\let\oldxLambda=\Lambda \renewcommand{\Lambda}{{\mbs{\oldxLambda}}}
\let\oldOmega=\Omega \renewcommand{\Omega}{\mathit{\oldOmega}}
\let\oldxGamma=\Gamma \renewcommand{\Gamma}{{\mbs{\oldxGamma}}}
\let\oldvphi=\varphi \renewcommand{\varphi}{{\mbs{\oldvphi}}}
\renewcommand{\hat}{\widehat}
\newcommand{\stensors}{M^s}
\newtheorem{theorem}{Theorem}[section]


\begin{abstract}
\myabstract
\end{abstract}

\section{Introduction}
\label{sec-intro}
In linear and nonlinear structural theories (including bars, beams, plates, membranes, shells,
etc.), the kinematic definitions and the equilibrium equations need to be closed with
a relation between the strain measures and the stress resultants, possibly with time-dependent
and dissipative effects. In the simplest situations, often restricted to the linear elastic case, there exists
an analytical expression that relates stresses and strain resultants, leading to well-known
equations that have been employed for decades in structural analyses (see
\cite{Popov:1968va,Timoshenko:1972ue,Govindjee:2012be}, for example, among
many books fully describing this process).

Structural analyses involving nonlinear elastic or inelastic materials can rarely make use of
section response laws and must, therefore, relie directly on three-dimensional constitutive laws
evaluated at every point of the cross section. Such an evaluation is complicated due to the fact
that every structural theory constrains, from its outset, the kind of strains and stresses
that are allowed on the sections. The hypotheses imply that the constitutive response of
these material points is constrained and its evaluation is rarely possible by analytical means.

These constrained constitutive laws must provide, given a proper subset of the strain components and
knowing that some stress components vanish, the remaining parts of the strain and stress tensors. Numerical
methods designed to solve this problem are iterative by nature, since, except for the linear elastic
case, the strain-stress relations are often implicit and nonlinear. Standard Newton-Raphson
iterative methods -- or related ones -- have been proposed for this purpose in the past
\cite{deBorst:1991ta,klinkel_using_2002,Valoroso:2009kh,deSouzaNeto:2011ve}.

Newton-like methods solve the constrained constitutive problem in a fast and robust fashion. By
focusing on the nonlinear (and constrained) strain-stress relations, however, they fail to notice
that there is an underlying variational structure behind the problem. In this article we reveal the
precise meaning of this framework, and identify the form of the energy functional involved. Such a
characterization allows to show that the equilibrium problem of structures, even when the section
constitutive law is evaluated pointwise from contrained three-dimensional models, has a global
variational character.

This assertion has at least three consequences. First, from the theoretical
point of view, theorems for the existence of solutions can now be derived based on minimizing
sequences and properties of functionals, even in the constrained case. Second, for more pragmatic
reasons, numerical optimization methods can be used to solve the global problem \emph{and} local
constitutive relations, for example of nonlinear conjugate gradient type. And third, since a single
energy functional is shown to be the quantity to minimize for a given solution, it provides a
natural error indicator for Galerkin-type approximations.
The first of these three implications falls outside the scope of the current work,
but the second and the third point are
discussed and exemplified in Sections~\ref{sec-numerical-solution} and~\ref{sec-numerical-examples}.

The setting employed for the derivations in this article is that of linearized kinematics and
nonlinear materials with internal variables. This framework is general enough to accommodate many
problems of interest, and can be extended, in a fairly straightforward manner to include geometrical
nonlinearities. Due to the similarities with the small strain case, the details of such
generalization are not worked out in this article, although both small and finite strain simulation
examples will be provided.

The way that we choose to deal with material nonlinearities, rate and history
effects is by means of incremental potentials \cite{Ortiz:1999tq}. With them, a variational
structure is preserved even for problems involving plasticity, viscoelasticity, etc., as long as
this is interpreted in an incremental fashion. Such ideas, widespread in Computational Mechanics,
are extended by our results to structural models of inelastic materials.

The remainder of this article is organized in the following way. In Section~\ref{sec-structure} we
describe structural models in an abstract way that can encompass all types. In this completely
general setting, we introduce the connection between the section response and the constrained material
constitutive law. The underlying variational setting is discussed in
Section~\ref{sec-variational} by revealing that there is a compatible minimization structure at the
level of material point, section, and structural member. In addition to the
theoretical advantages of the formulation, as already mentioned, some numerical implications
are studied in Section~\ref{sec-numerical-solution} and illustrated by means of examples in
Section~\ref{sec-numerical-examples}. Section~\ref{sec-summary} closes the article summarizing the
main findings.

As will be detailed in Section~\ref{sec-numerical-solution}, we classify two types of methods that
can be employed to solve numerically the constrained or reduced constitutive model. The material open-source
library MUESLI \cite{portillo_muesli_2017} implements both, and readers can access it to verify the
details of the numerical implementation.

\section{From structural models to 3D, and back}
\label{sec-structure}
We explore in this section the relationship between the mechanics of three-dimensional solids and
structural theories. There are many interesting points of view for this analysis, including
asymptotic behavior, model reduction, numerical discretization, etc., but we restrict our exposition
to those features that help to explain the central point of this article: how three-dimensional
material constitutive laws can be employed to formulate section response relations. We address fully
inelastic materials, although we restrict our exposition to linearized kinematics. Extension to
the geometrically nonlinear regime follows in a relatively straightforward manner and, in fact,
we will employ it in some of the simulations of Section~\ref{sec-numerical-examples}.

\subsection{Abstract structural models}
\label{subs-abstract}
To accommodate all potential structural models in our analysis we present them next in an abstract
way. The notation is general enough to encompass bars, rods, plates, etc. and any other
model based on structural strain measures and stress resultants. When possible, we will
make the connection to specific models, but the goal is to set up a common framework to
establish the relation with three-dimensional material models in a single form. We proceed
in this line by introducing six crucial elements of every structural theory:

\paragraph{i) Geometry} The reference configuration of any structural model can be defined
geometrically as a product bundle $\mathcal{S}=\mathcal{M}\times\mathcal{F}$, where the set $\mathcal{M}$
is referred to as the \emph{base} manifold and $\mathcal{F}$ is the \emph{fiber}
\cite{Epstein:2010vi}. Both sets are assumed to be smooth manifolds endowed with measures
$\mu(\mathcal{M})$ and $\mu(\mathcal{F})$, respectively. Key to any product manifold, there exist
surjective projections $\pi_M:\mathcal{S}\to\mathcal{M}$ and $\pi_F:\mathcal{S}\to\mathcal{F}$. Any
point on the model has coordinates $(\mbs{x},\mbs{\xi})$ with $\mbs{x}\in\mathcal{M}$ and
$\mbs{\xi}\in\mathcal{F}$. For example, if $\mathcal{S}$ is a shell, the base manifold is a
two-dimensional (smooth) surface that corresponds to the shell midsurface and the fiber is a segment
through the thickness.  When $\mathcal{S}$ is a rod, the base manifold is a curve and the fiber is
now the cross section.

\paragraph{ii) Generalized deformation} As in every mechanical theory, structural models
start from the definition of a configuration space $Q$ and a deformation
$\mbs{\chi}:\mathcal{M}\to Q$. The configuration space might include the displacements,
rotations, drill-free rotations, etc., defined at each point of the base manifold
and thus, in general, it might not even be a linear space. There
must be a natural embedding of every placement $\mbs{\chi}(\mathcal{M})$ on $\mathbb{R}^d$,
where $d=1,2,$ or~$3$.

\paragraph{iii) Strain measures} Depending on the particular geometry of the structural model
and the kind of relevant deformations, a set of characteristic strain measures
$\mbs{\Omega}=\hat{\mbs{\Omega}}(\mbs{\chi}(\mbs{x}))$ must be introduced. This set
must be frame invariant and gauge the relative changes in length and angle. For shell models,
for instance, the strain measure must include midsurface distortions,
transverse shear, and curvatures. Similarly, for a rod model, the strain measure
should include axial and shear deformations, as well as bending and torsional strains.

\paragraph{iv) Stored energy} An \emph{elastic} structural model assumes the existence
of a fiber stored energy density $U = \hat{U}(\mbs{\Omega};\mbs{x})$ such that the
total potential energy of the structure is
\begin{equation}
  V(\mbs{\chi}) :=
  \int_{\mathcal{M}} \hat{U}(\mbs{\Omega}(\mbs{\chi}(\mbs{x})); \mbs{x})\; \mu(\mathcal{M})
  + V_{ext}(\mbs{\chi})\ ,
  \label{eq-potential-energy}
\end{equation}
where $V_{ext}$ is the potential energy of the external forces. The explicit dependency
of $\hat{U}$ on $\mbs{x}$ indicates that the fiber response might be inhomogeneous. 

The constitutive modeling
of the \emph{inelastic} response is more complex. The fiber stored energy function, in these
cases, depends also on a set of frame-invariant fiber internal variables $\mbs{\alpha}$ so
that $U=\hat{U}(\mbs{\Omega}, \mbs{\alpha}; \mbs{x})$ and a supplemental kinetic
relation must be provided to model their evolution. The most common case is when
a kinetic potential $\hat{\psi} $ exists, so that
\begin{equation}
  \dot{\mbs{\alpha}} = \pd{\hat{\psi}}{\mbs{Q}}(\mbs{Q};\mbs{x}),
  \qquad
  \mathrm{with}
  \qquad
  \mbs{Q} := - \pd{U}{\alpha}(\mbs{\Omega}, \mbs{\alpha}; \mbs{x}) \ .
  \label{eq-kinetic}
\end{equation}
Again, the kinetic potential might be inhomogeneous and hence depend explicitly on
the base point~$\mbs{x}$. Often, it is required that $\hat{\psi}$ is convex,
although not necessarily differentiable, since this is enough to guarantee
non-negative dissipation.

\paragraph{v) Stress resultants.} Work conjugate to the strain measures,
the stress resultant $\mbs{\Sigma}$ is introduced. For either elastic
or inelastic section response, the thermodynamic definition of
this resultant is
\begin{equation}
  \mbs{\Sigma} =
  \hat{\mbs{\Sigma}}
  (\mbs{\Omega}, \mbs{\alpha}; \mbs{x})
  :=
  \pd{\hat{U}}{\mbs{\Omega}}
  (\mbs{\Omega}, \mbs{\alpha}; \mbs{x})
  \ .
  \label{eq-stress-resultant}
\end{equation}
In the usual structural models, $\mbs{\Sigma}$ includes the axial and shear
forces, and all the moments exerted on the fiber.

\paragraph{vi) Tangent operator.} The linearized tensor of elasticitiy is
often required in practical calculations. This operator is defined
as
\begin{equation}
  \mathbb{C} := \pd{\mbs{\Sigma}}{\mbs{\Omega}}\ .
  \label{eq-tangent}
\end{equation}
Its computation for elastic sections is usually simple, although its
extension to nonlinear ones needs to take into consideration the evolution
of the internal variables.

\subsection{Three-dimensional material response}
\label{subs-3d}
A closed form expression of the fiber stored energy density function $\hat{U}$ is only
known for the simplest cases, and in many problems of interest one must
rely on the general three-dimensional theory to calculate the section
response. For this reason we briefly summarize the key concepts in
the modeling of materials with internal variables, and then its restriction
to constrained response.

The mechanical state of materials with internal variables is fully described
by the strain $\mbs{\varepsilon}$ and a collection of internal variables $\mbs{\beta}$
that model all the inelastic phenomena. Then, the existence of a, possibly inhomogeneous,
stored energy function $W=\hat{W}(\mbs{\varepsilon},\mbs{\beta};\mbs{x},\mbs{\xi})$ is
assumed such that the stress tensor $\mbs{\sigma}$ is given by the derivative
\begin{equation}
  \mbs{\sigma} = \pd{\hat{W}}{\mbs{\varepsilon}}(\mbs{\varepsilon},\mbs{\beta};\mbs{x},\mbs{\xi})\ .
  \label{eq-sigma}
\end{equation}
An evolution equation for the internal variables must be supplied
with a kinetic potential $\hat{\phi}$, for example in the
form of
\begin{equation}
  \dot{\mbs{\beta}} = \pd{\hat{\phi}}{\mbs{q}}(\mbs{q};\mbs{x},\mbs{\xi}) ,
  \qquad
  \mathrm{with}
  \qquad
  \mbs{q} :=  - \pd{\hat{W}}{\mbs{\beta}}(\mbs{\varepsilon},\mbs{\beta};\mbs{x},\mbs{\xi}) .
  \label{eq-3d-kine}
\end{equation}
As indicated for the fiber response, the kinetic potential is often selected to be
convex in order to guarantee a non-negative dissipation in every possible process.

\subsection{Linking three-dimensional and structural models}
\label{subs-link}
The connection between fiber and three-dimensional material models depends on a kinematic and a
kinetic hypotheses. The first one imposes that, given the value of the strain measure $\mbs{\Omega}$
at a point $\mbs{x}\in\mathcal{M}$, a part of the strain tensor~$\mbs{\varepsilon}$ on the fiber
$\pi_M^{-1}(\mbs{x})$ is known. More specifically, let  $\mathbb{S}^d$ be the set of symmetric,
linear operators from $\mathbb{R}^d$ to $\mathbb{R}^d$ and consider a split
$\mathbb{S}^d=\first{\mathbb{S}}\times\second{\mathbb{S}}$ with $0 < \dim{\first{\mathbb{S}}} = d
(d+1)/2 - \dim{\second{\mathbb{S}}} < d (d+1)/2$. Given that $\mbs{\varepsilon}=
(\first{\mbs{\varepsilon}}, \second{\mbs{\varepsilon}})$, the \emph{kinematic hypothesis} of a
structural model is a functional expression
\begin{equation}
  \first{\mbs{\varepsilon}}(\mbs{x},\mbs{\xi})
  =
  \hat{\first{\mbs{\varepsilon}}}(\mbs{\Omega}(\mbs{x}), \mbs{\xi}) .
  \label{eq-eps-constraint}
\end{equation}
This relation
expresses the fact that $\first{\mbs{\varepsilon}}$ is completely known
from the strain resultant $\mbs{\Omega}$ at the base point $\mbs{x}$ and
the position $\mbs{\xi}$ on the fiber. The specific form of such a relation
depends, naturally, on the particular model under consideration.

Since the stress tensor $\mbs{\sigma}$ also belongs to $\mathbb{S}^d$, it admits the split
$\mbs{\sigma}=(\first{\mbs{\sigma}}, \second{\mbs{\sigma}}) \in \first{\mathbb{S}}\times
\second{\mathbb{S}}$. The two parts of the stress are defined as
\begin{equation}
  \first{\mbs{\sigma}}
  \defined
  \pd{\hat{W}}{\first{\mbs{\varepsilon}}}
  (\mbs{\varepsilon},\mbs{\beta};\mbs{x},\mbs{\xi})
  \ ,
  \qquad
  \second{\mbs{\sigma}}
  \defined
  \pd{\hat{W}}{\second{\mbs{\varepsilon}}}
  (\mbs{\varepsilon},\mbs{\beta};\mbs{x},\mbs{\xi})\ .
  \label{eq-split-dw}
\end{equation}
The \emph{kinetic hypothesis} of a structural model is a constraint
of the form
\begin{equation}
  \second{\mbs{\sigma}} = \mbs{0} ,
  \label{eq-sigma-constraint}
\end{equation}
that sets to zero precisely the part of the stress in the
space where the strains are not determined by the kinematic hypothesis.

The three-dimensional constitutive equations~\eqref{eq-3d-kine} and~\eqref{eq-split-dw}, together
with the constraint~\eqref{eq-sigma-constraint}, set up an implicit map $\first{\mbs{\sigma}} =
\hat{\first{\mbs{\sigma}}}(\first{\mbs{\varepsilon}},\mbs{\beta};\mbs{\xi})$, which we refer to as a
\emph{reduced constitutive model}.  Once this relation is established, its tangent relation
\begin{equation}
  \mathsf{C} 
  \defined
  \dd{\first{\mbs{\sigma}}}{\first{\mbs{\varepsilon}}}
  \label{eq-reduced-c}
\end{equation}
can be obtained.

In a structural model, the stress resultant $\mbs{\Sigma}$ at the base point $\mbs{x}$ can be
obtained from the integral over the fiber $\pi_M^{-1}(\mbs{x})$ of certain functions depending on
the fiber coordinate~$\mbs{\xi}$ and the reduced stress $\first{\mbs{\sigma}}$.  Denoting this
relation as $f$, but leaving it undefined for the moment, one should be able to write
\begin{equation}
  \mbs{\Sigma}
  =
  {\mbs{\Sigma}}(\mbs{\Omega}, \mbs{\alpha}; \mbs{x})
  =
  \int_{\pi_{M}^{-1}(\mbs{x})}
  f(
  \first{\mbs{\sigma}}
  (\first{\mbs{\varepsilon}}, \mbs{\beta}; \mbs{x},\mbs{\xi}))
  \; \mu(\mathcal{F})\ .
  \label{eq-resultant-integral}
\end{equation}

\begin{figure}[t]
  \begin{centering}
      \includegraphics[width=1.0\textwidth]{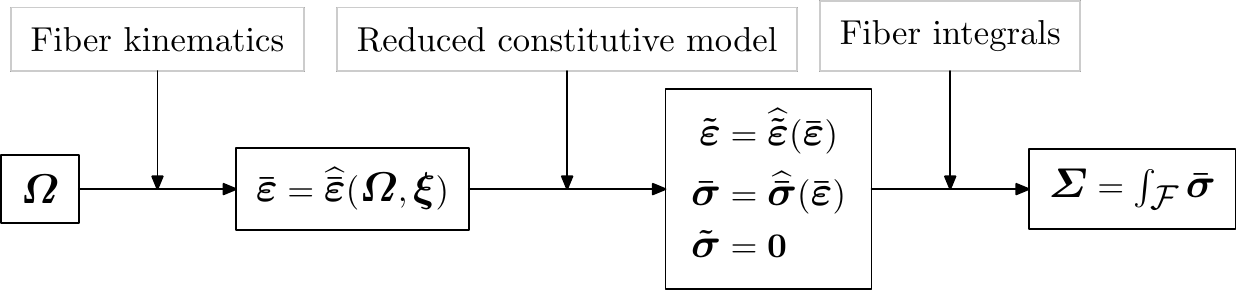}
      \caption{Logical structure behind the link of three-dimensional material
      laws and fiber response.}
  \label{fig-3-to-s}
  \end{centering}
\end{figure} 

Figure~\ref{fig-3-to-s} reveals the structure in the connection between structural and
three-dimensional constitutive models, as explained above, and whose
summary is as follows: in a structural model, when the strain measure
$\mbs{\Omega}$ is given at a point $\mbs{x}\in\mathcal{M}$, the fiber
kinematics assumption~\eqref{eq-eps-constraint} provides the value
of $\first{\mbs{\varepsilon}}$ at each point of the fiber $\pi_M^{-1}(\mbs{x})$.
With this information, and using the kinetic constraint~\eqref{eq-sigma-constraint},
the reduced constitutive law provides the stress $\first{\mbs{\sigma}}$. Finally,
the stress resultant can be calculated using Eq.~\eqref{eq-resultant-integral}.

\section{Variational structure}
\label{sec-variational}
We show in this section that the constitutive relation of a structural section
has a variational structure that encompasses the kinematic and kinetic hypotheses.
Such a property allows the formulation of the whole quasistatic problem
of equilibrium as a structural problem in the form of a minimization problem,
including the section response derived from three-dimensional constitutive theory.
When the material response is inelastic, however, such variational statement
can only be made \emph{incrementally}, which in fact is advantageous when searching
for numerical approximations. This characterization reveals that descent-type
numerical methods can be employed for the solution of general type of structural problems
and that the Hessian of the problem must always be symmetric, even in the presence of constraints.

To clearly identify the variational structure behind complex structural problems, we
analyze independently the problem at the level of points, fibers, and structure. We show next
that we can precisely characterize the variational principle behind each step.

\subsection{Variational update at the level of three-dimensional constitutive law}
\label{subs-var-point}
A material point with a constitutive relation, be it elastic or inelastic, as
described in Section~\ref{subs-3d}, can often be associated with an incremental
potential that defines its response. To see this, consider first that the
solution is obtained incrementally: given the strain $\mbs{\varepsilon}_n$
and the internal variables $\mbs{\beta}_n$, respectively, at time $t_n$, and the
strain $\mbs{\varepsilon}_{n+1}$ at time $t_{n+1} = t_n+\Delta t$,
the stress at this instant can be obtained as
\begin{equation}
  \mbs{\sigma}_{n+1} = \pd{W_n}{\mbs{\varepsilon}_{n+1}}(\mbs{\varepsilon}_{n+1}; \mbs{x},
  \mbs{\xi}) ,
  \label{eq-point-wn}
\end{equation}
where $W_n$ is an \emph{effective} stored energy function that depends on the state
$(\mbs{\varepsilon}_n,\mbs{\beta}_n)$. This kind of effective potentials, giving rise to so-called
\emph{variational updates}, are standard, and we refer to existing articles for their precise
definition (cf. \cite{Radovitzky:1999kc,Ortiz:1999tq,Miehe:2002je,Mosler:2010tv}, among others).

\subsection{Variational form of the reduced constitutive model}
\label{subs-var-reduced}
In Section~\ref{subs-link}, we defined a \emph{reduced constitutive model} as one relating the
strains~$\first{\mbs{\varepsilon}}$, the internal variables~$\mbs{\beta}$, and the
stress~$\first{\mbs{\sigma}}$, that satisfies the constraint $\second{\mbs{\sigma}}= \mbs{0}$. We
show next that for any three-dimensional material model, and any value of
$\first{\mbs{\varepsilon}}$ and $\second{\mbs{\sigma}}$ (the latter not necessarily equal to zero),
we can describe the problem of finding $\first{\mbs{\sigma}}$ and
$\second{\mbs{\varepsilon}}$ within a variational framework.

In what follows, we consider elastic and inelastic material models and we assume that the
constitutive model has an (incrementally) variational basis, as described in
Section~\ref{subs-var-point}. Within this setting, the key aspect of the reduced model is defining a
map
\begin{equation}
  (\first{\mbs{\varepsilon}}_{n+1}, \second{\mbs{\sigma}}_{n+1})\mapsto
  \second{\mbs{\varepsilon}}_{n+1}  
  \label{eq-red-update-map}
\end{equation}
that is consistent with the three-dimensional constitutive
law. Once the full strain $(\first{\mbs{\varepsilon}}_{n+1}, \second{\mbs{\varepsilon}}_{n+1})$
is known, the calculation of the
stored energy, stress, and tangent elasticities is trivial. The sought map is characterized
variationally by the optimization problem
\begin{equation}
  \second{\mbs{\varepsilon}}_{n+1}
  =
  \arg\sup_{\mbs{\gamma}}
  \left( \mbs{\gamma}\cdot \second{\mbs{\sigma}}_{n+1} -
    W_n( \first{\mbs{\varepsilon}}_{n+1}, \mbs{\gamma};
    \mbs{x},
    \mbs{\xi})
  \right)\ ,
  \label{eq-optim-reduced}
\end{equation}
which has a unique solution, provided $W_n$ is a convex function of its second argument.
Assuming $W_n$ is differentiable, the solution to Eq.~\eqref{eq-optim-reduced} is
dictated by the optimality condition
\begin{equation}
  0 = \second{\mbs{\sigma}}_{n+1} -
  \pd{W_n}{\second{\mbs{\varepsilon}}_{n+1}}(\first{\mbs{\varepsilon}}_{n+1},
    \second{\mbs{\varepsilon}}_{n+1};\mbs{x},\mbs{\xi})\ ,
  \label{eq-optim-reduced2}
\end{equation}
which is precisely the required implicit map~\eqref{eq-red-update-map}. Interestingly,
the function
\begin{equation}
  W_n^*(\first{\mbs{\varepsilon}}_{n+1}, \second{\mbs{\sigma}}_{n+1}; \mbs{x}, \mbs{\xi})
  :=
  \sup_{\mbs{\gamma}}
\left( \mbs{\gamma}\cdot \second{\mbs{\sigma}}_{n+1} -
    W_n( \first{\mbs{\varepsilon}}_{n+1}, \mbs{\gamma}; 
    \mbs{x},
    \mbs{\xi})
  \right)    
  \label{eq-red-legendre}
\end{equation}
is the Legendre transform of the incremental stored energy
function $W_n(\first{\mbs{\varepsilon}}, \second{\mbs{\varepsilon}}; \mbs{x}, \mbs{\xi})$
with respect to its second argument. Hence, by the properties of this
transform, the solution to the update map~\eqref{eq-red-update-map} can
be written in closed form as
\begin{equation}
  \second{\mbs{\varepsilon}}_{n+1}
  =
  \pd{ W_n^*}{\second{\mbs{\sigma}}_{n+1}}
  (\first{\mbs{\varepsilon}}_{n+1},\second{\mbs{\sigma}}_{n+1}; \mbs{x}, \mbs{\xi})\ .
  \label{eq-red-dual}
\end{equation}
A closed-form expression of $\second{\mbs{\varepsilon}}_{n+1}$ relies on
an explicit calculation of the dual $W_n^*$, which might
not be possible in many cases.

Irrespective of whether an analytical expression exists for $W_n^*$ or not, the
reduced constitutive relations can always be written as
\begin{equation}
  \first{\mbs{\sigma}}_{n+1}
  =
  \pd{W_n^*}{\first{\mbs{\varepsilon}}}
  (\first{\mbs{\varepsilon}}_{n+1},\second{\mbs{\sigma}}_{n+1}; \mbs{x}, \mbs{\xi})
  \ ,
  \qquad
  \mathsf{C}
  =
  \frac{\partial^2 W_n^*}{\partial \first{\mbs{\varepsilon}}^2}
  (\first{\mbs{\varepsilon}}_{n+1},\second{\mbs{\sigma}}_{n+1}; \mbs{x}, \mbs{\xi})
  .
  \label{eq-red-model}
\end{equation}
In these two expressions, the stress $\second{\mbs{\sigma}}_{n+1}$ plays the
role of a parameter and in all cases of interest it is equal to $\mbs{0}$. Hence,
and for future reference, we define the \emph{incremental reduced stored energy density}
to be
\begin{equation}
  \bar{W}_n(\first{\mbs{\varepsilon}}_{n+1}; \mbs{x}, \mbs{\xi})
  :=
  W_n^*(\first{\mbs{\varepsilon}}_{n+1}, \mbs{0}; \mbs{x}, \mbs{\xi})\ .
  \label{eq-effective-w}
\end{equation}

\subsection{Variational form of the section response}
The stored energy $U$ was defined in Section~\ref{sec-structure} for any
structural member as a function of the section strain $\mbs{\Omega}$
and the section internal variables $\mbs{\alpha}$. Linking
the section response with the three-dimensional material law, as
described in Section~\ref{subs-link}, the energy $U_n$ can 
be defined as
\begin{equation}
  \hat{U}_n(\mbs{\Omega}_{n+1}; \mbs{x})
  :=
  \int_{\mathcal{F}}
  \bar{W}_n(\first{\mbs{\varepsilon}}_{n+1}; \mbs{\xi})
  \; \mu(\mathcal{F})\ ,
  \label{eq-var-section}
\end{equation}
where $\first{\mbs{\varepsilon}}$ itself is a function of the
strain $\mbs{\Omega}$ and the fiber coordinate, as in Eq.~\eqref{eq-eps-constraint}.

The stress resultant $\mbs{\Sigma}$ can be obtained using Eq.~\eqref{eq-stress-resultant}
from the section potential, giving:
\begin{equation}
  \mbs{\Sigma}_{n+1}
  = \pd{ \hat{U}_n}{\mbs{\Omega}_{n+1}}
  =
  \int_{\mathcal{F}} \pd{\bar{W}_n}{\first{\mbs{\varepsilon}}_{n+1}}
  \cdot
  \pd{ \first{\mbs{\varepsilon}}_{n+1}}{\mbs{\Omega}_{n+1}}
  \; \mu(\mathcal{F})
  =
  \int_{\mathcal{F}}
  \first{\mbs{\sigma}}_{n+1} 
  \cdot
  \pd{ \first{\mbs{\varepsilon}}_{n+1}}{\mbs{\Omega}_{n+1}}
  \; \mu(\mathcal{F}) .
  \label{eq-var-sec-stress}
\end{equation}
In the previous identities we have employed the definition of the
stress presented in Eq.~\eqref{eq-red-model}. Comparing this
result with the definition~\eqref{eq-resultant-integral} we
identify the function $f$ in the latter equation with the relation
\begin{equation}
  f( \first{\mbs{\sigma}}_{n+1}; \mbs{\xi})
  =
  \first{\mbs{\sigma}}_{n+1} 
  \cdot
  \pd{ \first{\mbs{\varepsilon}}_{n+1}}{\mbs{\Omega}_{n+1}} .
  \label{eq-var-f-function}
\end{equation}

\subsection{Global variational problem}
\label{subs-global-var}
The previous results enable us to write the global equilibrium
problem of a structural model, including the reduced constitutive
model, as a single minimization problem. The generalized
displacement of the structure is the one solving the problem
\begin{equation}
  \inf V(\mbs{\chi})
  =
  \inf
  \int_{\mathcal{M}}
  \hat{U}_n(\mbs{\Omega}(\mbs{\chi}_{n+1}); \mbs{x})
  \;\mu(\mathcal{M})
  +
  V_{ext}(\mbs{\chi}_{n+1})\ ,
  \label{eq-global}
\end{equation}
with $U_n$ being the section incremental potential,
defined in Eq.~\eqref{eq-var-section}.
\section{Solution of the reduced constitutive equations}
\label{sec-numerical-solution}
The structural models described in this article can employ
arbitrary (small strain) material models making use of
incremental updates and setting the problem statement
under a single variational framework. Key to this unified
formulation is the identification of the reduced
effective stored energy function $\bar{W}_n$ that accounts
for the kinetic constraints.

As explained in Section~\ref{sec-structure}, the reduced
constitutive model consists of finding part of the stress
and the strain tensors at a point, given the remaining components
of these two objects. For inelastic problems written in
an incremental fashion, the crucial step is the determination
of $\second{\mbs{\varepsilon}}_{n+1}$ given $\first{\mbs{\varepsilon}}_{n+1}$
and the relation
\begin{equation}
  \mbs{0}
  =
  \pd{W_n}{\second{\mbs{\varepsilon}}_{n+1}}
  (\first{\mbs{\varepsilon}}_{n+1}, \second{\mbs{\varepsilon}}_{n+1}; \mbs{x}, \mbs{\xi})\ , 
  \label{eq-nl-solver}
\end{equation}
where the coordinates $\mbs{x}$ and $\mbs{\xi}$ are known.

The most obvious strategy to find $\second{\mbs{\varepsilon}}_{n+1}$ consists in solving directly
the (possibly nonlinear) Eq.~\eqref{eq-nl-solver}.  A Newton-Raphson or quasi-Newton scheme
seems the most appropriate route in this case, since the rate of convergence is high, as long as the
radius of convergence in this problem is large enough and the computation of the Hessian of $W_n$ is
not too cumbersome or computationally demanding.

A second option to obtain $\second{\mbs{\varepsilon}}_{n+1}$ exploits the
underlying variational nature behind Eq.~\eqref{eq-nl-solver}. As explained
in Section~\ref{subs-var-reduced}, the unknown part of the strain can
also be characterized by
\begin{equation}
  \second{\mbs{\varepsilon}}_{n+1}
  =
  \arg \sup_{\mbs{\gamma}} (- W_n(\first{\mbs{\varepsilon}}, \mbs{\gamma}; \mbs{x}, \mbs{\xi}))
  =
  -
  \arg \inf_{\mbs{\gamma}} W_n(\first{\mbs{\varepsilon}}, \mbs{\gamma}; \mbs{x}, \mbs{\xi})\ .
  \label{eq-solu-mini}
\end{equation}
This expression indicates that the strain $\second{\mbs{\varepsilon}}_{n+1}$ can also be
found by minimizing the incremental stored energy function with respect to some
of its arguments. Hence, descent methods, such as the nonlinear conjugate gradient,
can be used to solve the reduced constitutive model in an effective way, without
the need for computing the Hessian, and with guaranteed convergence if $W_n$ is convex.

Both approaches for the solution of the reduced constitutive law, namely the Newton-Raphson
and the descent methods, are available, for a large number of common material laws, in the
open-source library MUESLI \cite{portillo_muesli_2017}.

\section{Numerical examples}
\label{sec-numerical-examples}
We examine next several numerical examples involving bars, beams, and shells combined with various
three-dimensional constitutive laws. The purpose of these simulations is, first, to
demonstrate that the two solution strategies outlined in Section~\ref{sec-numerical-solution}
are both valid routes for the integration of the section response for complex material
models. Second, we will show that the variational structure behind the constitutive
response of reduced models opens the door to error indicators.

In all problems the material library MUESLI \cite{portillo_muesli_2017} is used for computation
of the constitutive response of the corresponding reduced model. In the case of bars, the condition
$\second{\mbs{\sigma}} = (S^{12}, S^{13}, S^{22}, S^{23}, S^{33}) = \mbs{0}$ is considered by the solution
algorithms presented in Section~\ref{sec-numerical-solution}. In the case of beams, the constraint
take the form $\second{\mbs{\sigma}}=(S^{22}, S^{23}, S^{33})=\mbs{0}$. In the case of shells, the condition
$\second{\mbs{\sigma}}=(S^{33})=\mbs{0}$ is incorporated, where $S^{33}$ is the transverse normal stress
component of the second Piola-Kirchhoff stress tensor $\mbs{S}$.

\subsection{Bar problem}
\label{sec-bar}

\begin{figure}[t]
\centering
\includegraphics[width=.4\textwidth]{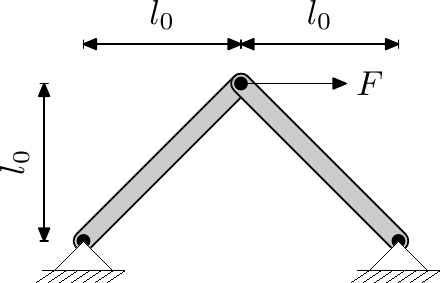}
\caption{Geometry and load of the bar problem.}
\label{fig-bar-problem}
\end{figure}

\begin{figure}[t]
\centering
\includegraphics[width=.75\textwidth]{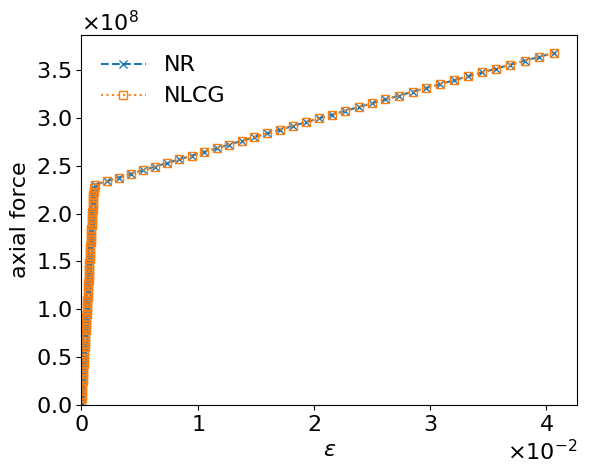}
\caption{Axial force vs elongation in the bar under tension.}
\label{fig:bar_sol}
\end{figure}

This first example consists of a simple plane truss under finite strain hypothesis. The structure
consists of two connected bars of equal length $l_0 = \sqrt{2}$ and section $A=10^{-2}$, pinned at
their ends (see Fig.~\ref{fig-bar-problem}). The pinned nodes have Cartesian coordinates $\left( 0.0,0.0 \right)$ and
$\left( 2.0, 0.0 \right)
$. In the central node, a horizontal force $F=5\cdot10^6$ is applied.  The material of the bars is
elasto-plastic with Young's modulus $E=210 \cdot 10^9$, Poisson's ratio $\nu = 0.3$, a von Mises
yield function with yield stress $\sigma_e =230 \cdot 10^6 $ and isotropic hardening modulus
$H=4\cdot10^9$.  Since the deformation of the cross section is uniform, only one quadrature point is
chosen on it to evaluate the section response.

This simple problem allows to demonstrate the use of a non-linear conjugate gradient (NLCG)
method to solve both the global problem and the local constitutive relations, and compare it
with the solution obtained with a Newton-Raphson solver, both at the global and local levels.
Figure~\ref{fig:bar_sol} shows the axial force vs elongation, $\epsilon = \frac{l-l_0}{l_0}$, in the
bar that is under tension. As expected, both methods give the same result, and we note that
the NLCG never requires the computation of a tangent operator.

\begin{figure}[t]
  \begin{centering}
      \includegraphics[width=0.4\textwidth]{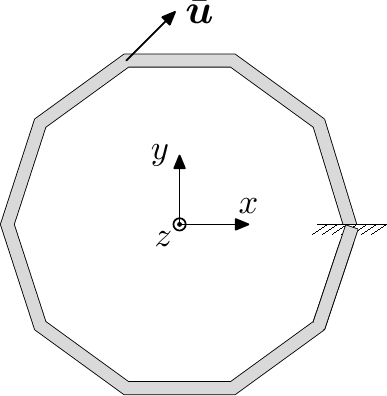}
  \caption{Geometry of beam problem.}
  \label{fig-beam-problem}
  \end{centering}
\end{figure} 

\subsection{Beam problem}
\label{sec-beam}

In the second example we study the response of a beam with the shape of a decagon, see
Figure~\ref{fig-beam-problem}, under the small strain hypothesis.  The beam axis lies in the $xy$-plane
and the circumscribed circle radius is $r$. The cross section at one of the nodes is clamped.  The
cross section at the third vertex of the decagon, counting form the clamped one, is subject to an
imposed displacement $\mbs{u}= \delta (\mbs{i} + \mbs{j} + \mbs{k})$, where the vectors
$(\mbs{i},\mbs{j},\mbs{k})$ refer to the canonical basis. The cross section of the beam is
rectangular with sides $a$ and $b$. The side with dimension $a$ is parallel to the $xy$ plane, and
the side with dimension $b$ is perpendicular to the first one and to the tangent direction of the
beam.  The material of the beam is elasto-plastic with Young's modulus $E$, Poisson's ratio $\nu$,
a von Mises yield function of yield stress $\sigma_e$ and isotropic hardening modulus $H$. Under
these conditions, the beam is subjected to axial, shear, bending, and torsion deformations.

\begin{figure}
\centering
\includegraphics[width=.3\textwidth]{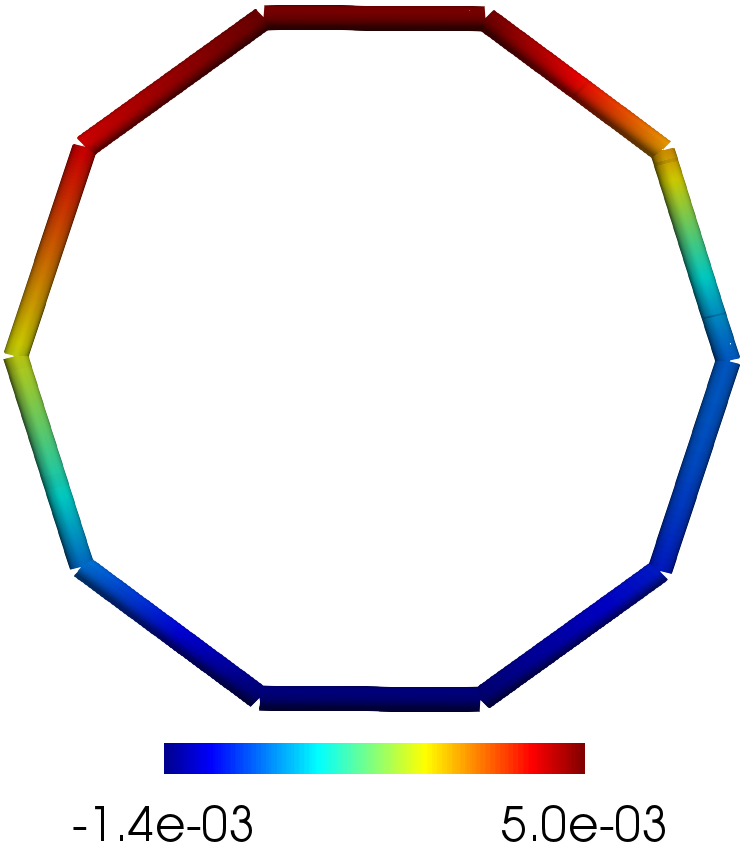}
\includegraphics[width=.3\textwidth]{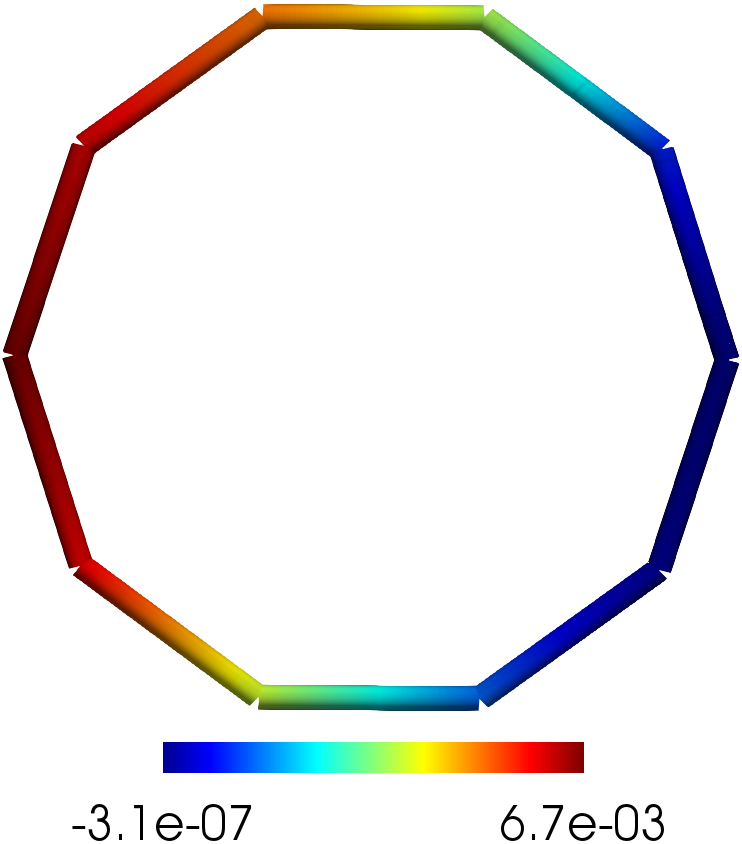}
\includegraphics[width=.3\textwidth]{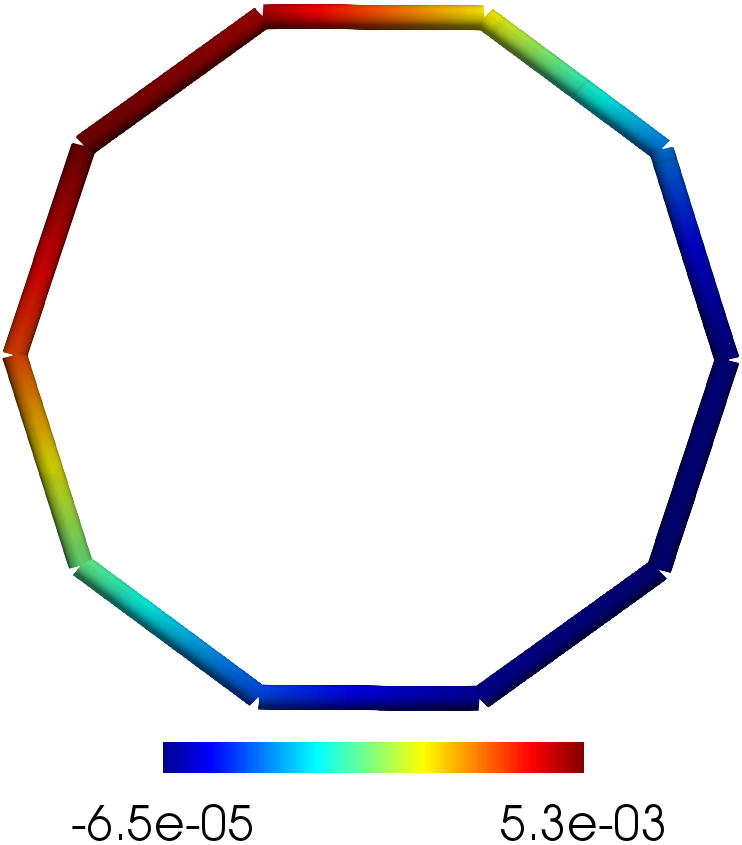}
\caption{Displacement in $x$ (left), $y$ (middle) and $z$ (right) for the beam problem at the
end of the simulation.}
\label{fig:beam_sol}
\end{figure}

For the numerical example we select $r = 0.5,\ \delta = 5.0\cdot10^{-3},\ E = 210\cdot10^9,\ \nu =
0.3,\ \sigma_e = 200\cdot10^6,\ H = 10\cdot10^9$ and the beam is discretized with two-node elements
with independent displacement and rotation interpolation, and selective reduced integration for
the shear terms. An $8\times 8$ quadrature rule is chosen for the section constitutive response.
Figure~\ref{fig:beam_sol} shows the values of the
displacements on the deformed shape of the beam when a mesh of 10 elements is employed.

\begin{figure}
\centering
\includegraphics[width=.48\textwidth]{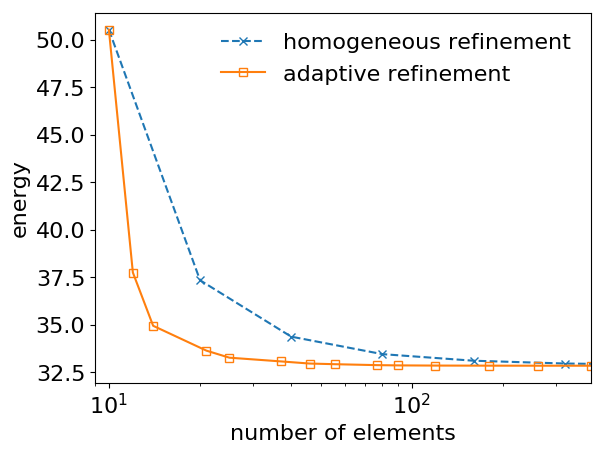}
\includegraphics[width=.48\textwidth]{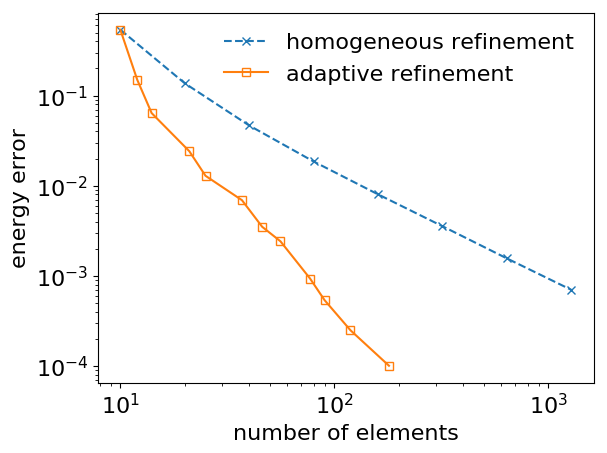}
\caption{Effective energy vs number of elements (left) and Relative energy error vs number of elements (right) for
  the homogeneous and adaptive mesh refinement strategies.}
\label{fig:beam_energy}
\end{figure}

In order to improve the solution, we refine the mesh homogeneously. Figure~\ref{fig:beam_energy}
(left) shows the value of the global effective potential as a function of the number of elements in
the mesh (in dashed blue line). As this number increases, the effective potential is reduced,
converging to an asymptotic value, the ``distance'' to which can serve as an indicator of the
accuracy of the solution.

\begin{figure}
\centering
\includegraphics[width=.48\textwidth]{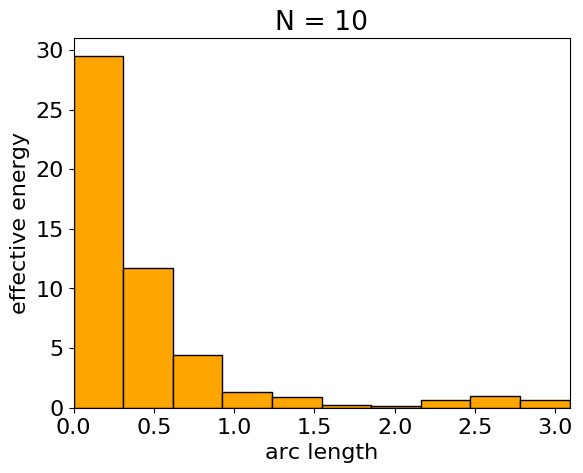}
\includegraphics[width=.48\textwidth]{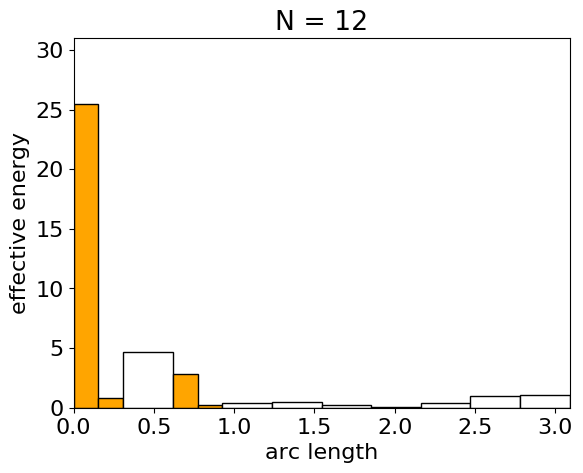}
\includegraphics[width=.48\textwidth]{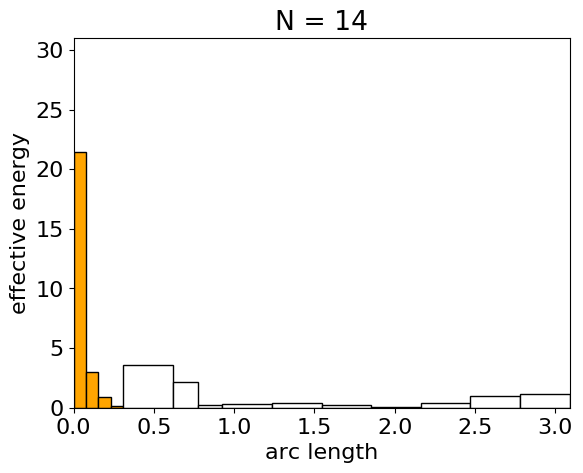}
\includegraphics[width=.48\textwidth]{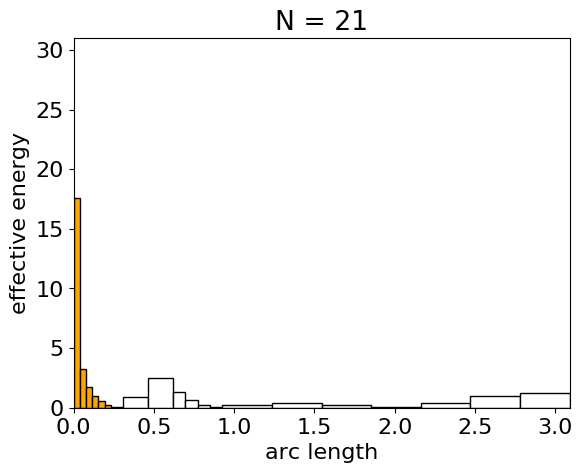}
\caption{Effective energy of each element for four consecutive iterations of the adaptive refinement strategy.
  In orange, the elements with the smallest length. The total number of elements in the mesh
is indicated above each plot.}
\label{fig:beam_energy_steps}
\end{figure}

The local value of the effective potential can be used as an indicator for local mesh refinement. In
a second solution, we proceed to refine the mesh by halving those beam elements that contribute most
to the global value of the effective potential. For that, we proceed as in \cite{Mosler:2011je},
studying the addition of a new node locally. For each element in the original mesh, we compare the
energy obtained in the latest refinement with the one obtained in a local problem that results from
halving the element and imposing the displacements and rotations of the latest calculation in the
boundary nodes. Then, we subdivide the element if this energy difference is larger than the global
average, hence searching for a more regular distribution of the effective potential energy density.
Figure~\ref{fig:beam_energy} (left)  depicts the value of the total effective energy as a function
of the number of elements refined in this anisotropic fashion (solid orange line). In Figure
~\ref{fig:beam_energy} (right) the relative energy error as a function of the number of elements for
both refinement strategies is depicted. When compared with the homogeneous refinement, it follows
that the locally refined mesh approximates faster the exact solution than the globally refined
one. This, in turn, justifies the consideration of the effective potential energy as a valid error
indicator. Figure~\ref{fig:beam_energy_steps} shows the efective potential energy per element in
four consecutive local refinements, with the element size can be deduced from the abscissae.

\subsection{Pinched cylinder}
\label{sec-cylinder}
\begin{figure}
      \centering\small
      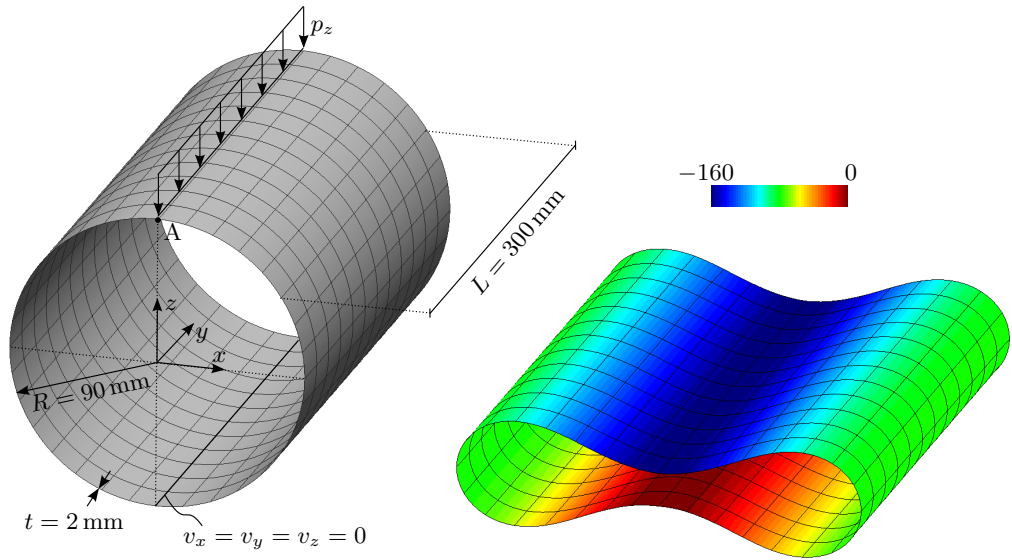
      \caption{Pinched cylinder, left: problem setup, right: deformed structure in final stage with corresponding
        contour plot of displacement $v_z$.}
 \label{fig:cylinder_setup}
\end{figure}

Next we consider two shell problems. Both are solved using a primal isogeometric Kirchhoff-Love
shell formulation with numerical integration through the shell's thickness, as presented in
\cite{oesterle_hierarchic_2017}. The shell formulation employed uses stresses, rather than stress
resultants, which is in contrast to most shell formulations found in literature. However, the
resulting formulation provides practically identical results as the formulations presented in
\cite{kiendl_isogeometric_2015} and \cite{ambati_isogeometric_2018}, where the initial isogeometric
Kirchhoff-Love shell formulation \cite{kiendl_isogeometric_2009}, which was based analytical
preintegration, was reformulated in terms of a numerical integration through the thickness of the
shell body.  All discretizations contain at least quadratic, $C^1$-continuous NURBS shape functions
within a patch. In the shell's in-plane directions, we use a ``full'' integration scheme with $p+1$
Gauss points, where $p$ is the polynomial degree of the shape functions. The number of integration
points in the thickness direction is variable, but chosen to be equal to three for the presented
examples.

As first shell problem, we study the pinching of a cylinder employing a hyperelastic material model,
i.\,e. a compressible Neo-Hooke material with Young's modulus of $E = 168 000 \UNpmmq$ and Poisson's
ratio of $\nu=0.4$. The problem setup for a very thick (slenderness $R/t=4.5$) and a moderately thin
cylinder (slenderness $R/t=45$) was first proposed by \cite{buchter_three-dimensional_1994} and
subsequently studied by variuos researchers, see for instance \cite{brank_nonlinear_2002,
schwarze_reduced_2011}. All the articles mentioned consider three-dimensional or solid shell
formulations making use of fully three-dimensional constitutive laws. Here, we only consider the
thin case, since a thin shell formulation of Kirchhoff-Love type is used, as it was done in
\cite{kiendl_isogeometric_2015}. For the sake of comparability, we use a discretization of the shell
problem in accordance with \cite{kiendl_isogeometric_2015}, i.\,e. we use a fourth order,
$C^3$-continuous discretization with $32\times12$ elements for modeling the whole cylinder, see
Figure~\ref{fig:cylinder_setup}. In \cite{kiendl_isogeometric_2015}, one half of the cylinder was
modeled with corresponding application of symmetry boundary conditions. With a slight difference, we
instead model the whole cylinder with four patches consisting of $8\times12$ elements each, coupled
with the bending strip method from \cite{kiendl_bending_2010}, where a penalty parameter has to be
chosen.

The simulation is performed by controlling the vertical displacement $v_{z,A}$ at point $A$ in 16
steps of $\Delta v_{z,A}=-10 \Umm$. For the maximum displacement
$v^{\text{max}}_{z,A}=-160 \Umm$ the resulting force $F^{\text{max}}$ is measured and compared to
values obtained in literature. For penalty parameters of $10^2$ and $10^3$ we obtain the resulting
forces $F^{\text{max}}=34.72 \UkN$ and $F^{\text{max}}=34.77 \UkN$, respectively. For higher penalty
parameters the convergence properties get worse. In \cite{kiendl_isogeometric_2015}, a value of
$F^{\text{max}}=34.86 \UkN$ was reported, whereas the results obtained with three-dimensional or solid
shell formulations vary from $F^{\text{max}}=34.59 \UkN$ to $F^{\text{max}}=35.47 \UkN$, as 
reported in \cite{schwarze_reduced_2011}. Our obtained results are consequently in good agreement
with the results from the literature.

\subsection{Simply supported plate}
\label{sec-plate}
\begin{figure}
      \centering\small
      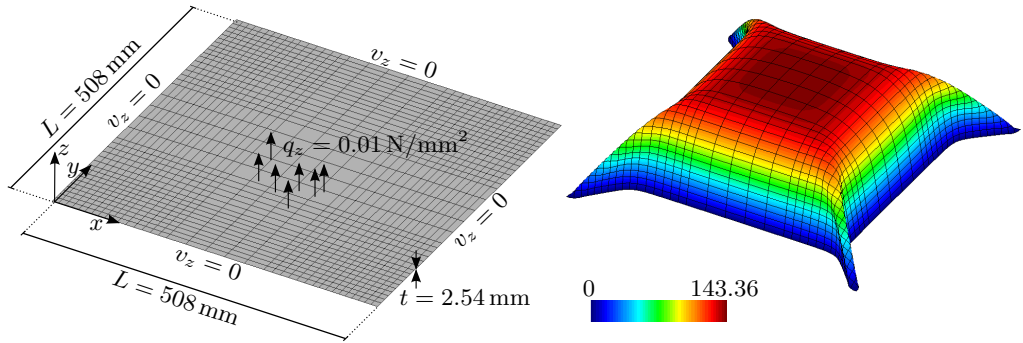
   \caption{Simply supported plate, left: problem setup, right: deformed structure in final stage with corresponding contour plot of displacement $v_z$.}
 \label{fig:plate_setup}
\end{figure}

The second shell problem consists of a simply supported plate with a large strain plastic material,
i.\,e. perfect $J_2$-plasticity with Young's modulus $E = 6.9\cdot 10^4 \UNpmmq$, Poisson's
ratio $\nu=0.3$ and yield stress $\si_y = 248 \UNpmmq$. As shown in
Figure~\ref{fig:plate_setup}, the quadratic plate has a side length of $L = 508 \Umm$ and a
thickness of $t=2.54 \Umm$, resulting in a slenderness ratio of $L/t=200$. All edges of the plate
are simply supported in $z$-direction and the plate is subjected to a uniform dead load of $q_z=0.01
\UNpmmq$ . The plate problem with this type of loading can be found in several references, see
for instance \cite{buchter_three-dimensional_1994, miehe_theoretical_1998,
hauptmann_`solid-shell_2001, reese_large_2007, klinkel_mixed_2008, schwarze_reduced_2011}.
\begin{figure}
      \centering\small
      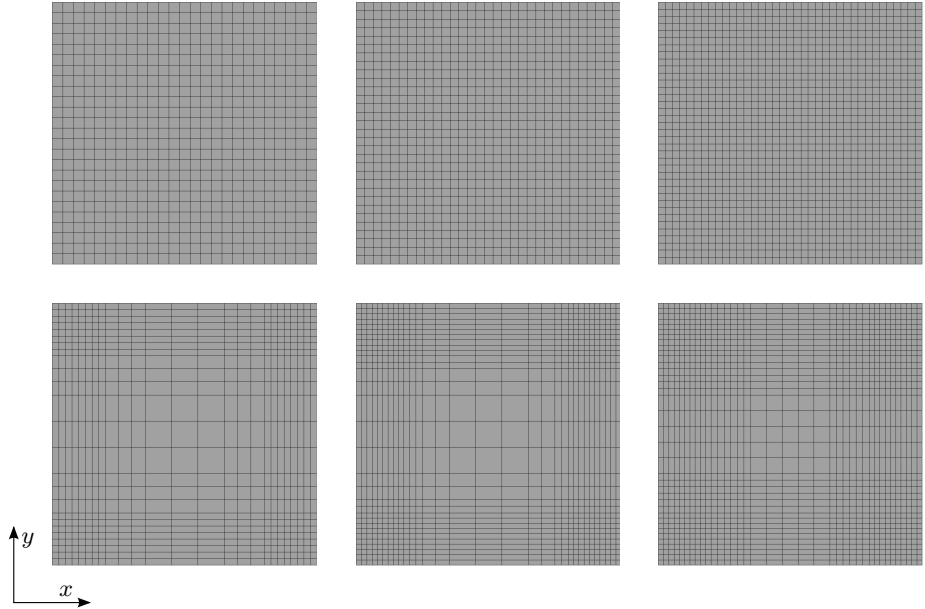
      \caption{Simply supported plate, employed meshes with quadratic, $C^1$-continuous B-splines, left: $25\times25$ elements,
        left: $31\times31$ elements, right: $37\times37$ elements, top: uniform meshes, bottom: nonuniform (NU) meshes.}
 \label{fig:plate_meshes}
\end{figure}
Although symmetry may be taken into account, we model the whole plate structure and
discretize it with six different, uniform and nonuniform (NU) refined meshes using quadratic,
$C^1$-continuous B-splines. The meshes employed are shown in Figure~\ref{fig:plate_meshes} and
consist of $25\times25$, $31\times31$, and $37\times37$ elements, respectively. In all cases, three
Gauss points through the thickness are used, since there is only a marginal change in the results
when changing the number of Gauss points in the thickness direction, for example to five. This
observation matches the one made in \cite{schwarze_reduced_2011}, where five Gauss points through
the thickness have been used for a trilinear solid shell finite element, although the results did
not sinificantly differ compared to the solution with two Gauss points in the shell's thickness
direction.

\begin{figure}
      \centering\small
      \input{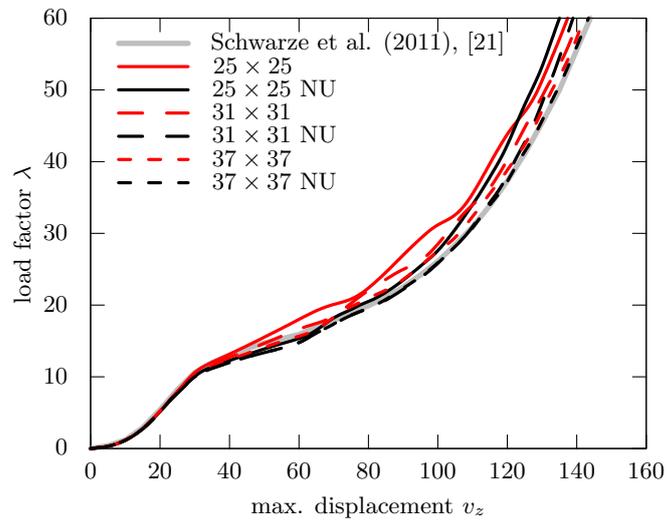}
   \caption{Simply supported plate, load-displacement diagram for different uniform and nonuniform (NU) meshes.}
 \label{fig:plate_ldk}
\end{figure}
\begin{figure}
      \centering\small
      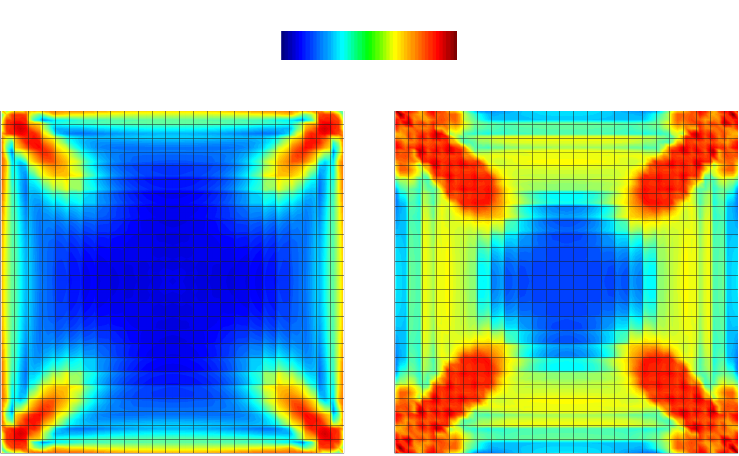
   \caption{Simply supported plate, effective stored energy density, integrated through the shell's thickness, with unit of $\mathrm{kNm}/\mathrm{m}^2$ at different load levels for uniform $25\times25$ element mesh, left: $\la=10$, right: $\la=20$.}
 \label{fig:plate_Energy}
\end{figure}

All simulations are performed with load control, using increments of different magnitude. We start
with ten increments of $\Delta \la= 0.2$, followed by ten increments of of $\Delta \la=
0.8$. Between $\la= 10$ and $\la= 20$, we use 50 increments of $\Delta \la= 0.2$, followed by 50
increments of $\Delta \la= 0.8$, until the maximum load level of 50 increments of $\la= 60$ is
reached.  More efficient time stepping schemes are not within the scope of this paper. The 
load-displacement curves obtained are plotted in Figure~\ref{fig:plate_ldk} and compared with the results
obtained in \cite{schwarze_reduced_2011} with a $64\times64$ uniform finite element mesh.  The
uniform $25\times25$ and $31\times31$ meshes behave too stiff in a large portion of the load-displacment
diagram, whereas the uniform $37\times37$ element mesh is only slightly too stiff. The effective
stored energy density distribution for the uniform $25\times25$ mesh is shown in
Figure~\ref{fig:plate_Energy} for two different load levels of $\la=10$ and $\la=20$. The plots
directly suggest a mesh refinement in the corner regions, for which the three NU meshes try to
account for.
The results for the $25\times25$ NU and $31\times31$ NU
meshes are slightly too stiff towards the maximum load level, probably because of a too coarse mesh size at the tip of the plastic fold lines.
The finest mesh with $37\times37$ elements produces results in good agreement with the results obtained in \cite{schwarze_reduced_2011}.
The curves practically match each other, except of a slight difference in the region between approximately $\la=10$ to $\la=15$.

\section{Summary and outlook}
\label{sec-summary}
Structural models employing complex material models (inelastic, rate-dependent, etc.)
require the solution of constrained
three-dimensional material constitutive laws for the evaluation of their section response. This
article starts by describing in an abstract way how these two problems are linked in
an arbitrary structural model.

The main result of this work is the identification of a variational setting behind
the problem of general structural models, encompassing three-dimensional
laws, both elastic and inelastic. Starting from a new result that identifies the
minimization problem behind every constrained three-dimensional constitutive relation,
and based on the abstract framework described first, we have revealed that there
is a single optimization program to be solved when analyzing general structural models
with complex materials. 

This result has several consequences: from the theoretical point of view it sets the stage for
fundamental analyses including the existence of solutions, an aspect that is beyond this work's
goals. From the computational side, it
guarantees the symmetry of the Hessian and points towards (descend) iterative methods for the
solution of these problems that have not been identified previously in the literature. Furthermore,
the effective potential that is the basis for the solution provides an intrinsic error indicator for
a general class of structural models, including geometric and material nonlinearities. This
is a remarkable result, and the first of its class to the authors' knowledge, illustrating
how the variational framework bridges the mechanics of three-dimensional and reduced models.

Numerical methods, implemented in a publicly available library of material models, show the
validity of the approach and illustrate the possibilities of mesh adaptation based on
the new error indicator.

\bibliographystyle{\mybibstyle}
\bibliography{literature}
\end{document}